\pdfoutput=1
\documentclass[11pt,a4paper]{article}
\usepackage{jheppub}
\usepackage{subfig}
\usepackage{graphicx}
\usepackage{hyperref}
\usepackage{multirow}
\usepackage{color}

\newcommand{\beq}{\begin{eqnarray}}
\newcommand{\eeq}{\end{eqnarray}}
\newcommand{\non}{\nonumber\\}

\newcommand{\p}{\partial}
\newcommand{\Tr}{\mathop{\rm Tr}}

\begin{document}

\title{A supersymmetric Skyrme model}

\author{Sven Bjarke Gudnason,${}^{1}$}
\author{Muneto Nitta${}^2$ and}
\author{Shin Sasaki${}^3$}
\affiliation{${}^1$Institute of Modern Physics, Chinese Academy of Sciences,
  Lanzhou 730000, China}
\affiliation{${}^2$Department of Physics, and Research and
    Education Center for Natural Sciences, Keio University, Hiyoshi
    4-1-1, Yokohama, Kanagawa 223-8521, Japan}
\affiliation{${}^3$Department of Physics, Kitasato University
Sagamihara 252-0373, Japan}
\emailAdd{bjarke(at)impcas.ac.cn}
\emailAdd{nitta(at)phys-h.keio.ac.jp}
\emailAdd{shin-s(at)kitasato-u.ac.jp}

\abstract{
Construction of a supersymmetric extension 
of the Skyrme term was a long-standing problem 
because of the auxiliary field problem; that is, 
the auxiliary field may propagate and cannot be eliminated,
and the problem of having fourth-order time derivative terms.
In this paper, we construct for the first time a supersymmetric
extension of the Skyrme term in four spacetime dimensions, 
in the manifestly supersymmetric superfield formalism 
that does not suffer from the auxiliary field problem.
Chiral symmetry breaking in supersymmetric theories 
results not only in Nambu-Goldstone (NG) bosons (pions)
but also in the same number of quasi-NG bosons 
so that the low-energy 
theory is described by an 
SL($N$,$\mathbb{C}$)-valued matrix field 
instead of SU($N$) for NG bosons.
The solution of auxiliary fields is trivial
on the canonical branch of the auxiliary field equation, 
in which case our model results in a fourth-order derivative term that
is not the Skyrme term.
For the case of SL(2,$\mathbb{C}$), 
we find explicitly a nontrivial solution to the algebraic auxiliary 
field equations that we call a non-canonical branch,
which when substituted back into the Lagrangian gives
a Skyrme-like model. If we restrict to a submanifold, where
quasi-NG bosons are turned off, which is tantamount to
restricting the Skyrme field to SU(2), then the fourth-order
derivative term reduces exactly to the standard Skyrme term. 
Our model is the first example of a nontrivial auxiliary field 
solution in a multi-component model.
}


\maketitle


\section{Introduction}

The Skyrme model was first introduced as a toy model describing
baryons in a low-energy mesonic field theory \cite{Skyrme:1962vh}. 
Later it was shown to be the low-energy limit of large-$N_c$
QCD \cite{Witten:1983tw}.
After this it gained popularity as a model of nuclei in the
literature.
It took, however, some time before the numerical calculations (and the
computing power) could tackle solutions for higher baryon numbers.
The breakthrough came with the introduction of the rational maps as an
approximation to the real Skyrmion solution \cite{Battye:1997qq}. 
These are very useful as initial guesses for numerical calculations. 
For vanishing pion mass, the fullerenes adequately described by the
rational maps are believed to be the global minimizers of the Skyrmion
energy functional.
Once a pion mass of the order of the physical pion mass is introduced,
the Skyrmions prefer to order themselves in a lattice of $B=4$ cubes,
which can be interpreted as a crystal of alpha
particles \cite{Battye:2006na}. 

Quite a few phenomenologically appealing results have been achieved in
the Skyrme model; for recent works, see e.g.~\cite{Lau:2014baa}. 
A withstanding problem of the Skyrme model, is that the binding
energies naturally come out too large (by about an order of
magnitude). 
For this reason, quite some work has been devoted to finding a BPS
limit of the Skyrme model.
The minimal (original) Skyrme model has a BPS
bound \cite{Faddeev:1976pg}, that, however, can be saturated only on 
the 3-sphere \cite{Manton:1986pz}. 
Recently, a different model has been suggested, called the BPS Skyrme
model \cite{Adam:2010fg}, which has a BPS limit and many exact
solutions have been found \cite{Adam:2015ele}. 
Naively, one may think that the BPS limits of the Skyrme model above
are related to supersymmetry, as is the case for
Abrikosov-Nielsen-Olesen vortices or for 't Hooft-Polyakov
monopoles \cite{Bogomolny:1975de}. 
This is, however, not the case for the Skyrme model.

There have been several attempts in the literature to construct 
supersymmetric extensions of the Skyrme model,
but all attempts failed and it turns out that 
this is indeed a very difficult problem.
The first attempt was made already three decades ago
by Bergshoeff-Nepomechie-Schnitzer \cite{Bergshoeff:1984wb}. 
The monkey-wrench is that the target space must be K\"ahler in order
for a supersymmetrization to be possible \cite{Zumino:1979et}.
$S^3$, the target space of the Skyrme model, is not K\"ahler (and
neither is $S^4$). 
Ref.~\cite{Bergshoeff:1984wb} chose to work around that obstacle by
gauging a U(1) subgroup of the right-acting SU(2) group. This gauging
reduces the target space to $\mathbb{C}P^1$, which is the poster boy
for K\"ahler spaces.
Their supersymmetrized model has, however, a typical feature of
supersymmetric higher-derivative models, namely that it possesses a
term with four time derivatives.
Or in other words, the Skyrme term is accompanied with 
an unwanted fourth-order time derivative term.
This in turn makes a Hamiltonian formulation impossible.
Essentially the same model was obtained by
Freyhult \cite{Freyhult:2003zb} in an attempt to supersymmetrize the
Faddeev-Skyrme model for Hopfions \cite{Faddeev:1996zj}.
More recently, one proposal was made in ref.~\cite{Queiruga:2015xka}, 
but it is not supersymmetric because a
non-supersymmetric constraint $|\phi_1|^2+|\phi_2|^2=1$, was put by
hand for the chiral superfields $\phi_{1,2}$.\footnote{The model may
  be made supersymmetric by the introduction of a vector multiplet,
  which would reduce the target space to $\mathbb{C}P^1$, similarly to
  Bergshoeff-Nepomechie-Schnitzer's approach, but for a sixth-order
  derivative term \cite{Queiruga:privatecommunication}. The model is thus
  a $\mathbb{C}P^1$-type model and the Skyrme model is not in its
  subspace. }

A more notorious problem called the auxiliary field problem 
arises when considering higher-derivative models with manifest
supersymmetry. 
The problem is that once derivatives act on the auxiliary field, its
equation of motion becomes dynamical instead of algebraic. This means
that the auxiliary field becomes propagating and cannot simply be 
eliminated.
This problem is related to the above mentioned problem and in fact was
encountered in both ref.~\cite{Bergshoeff:1984wb}
and \cite{Freyhult:2003zb}.  
Two situations occur. If the derivatives act on the auxiliary field
$F$ as $X\p F$, then the problem can be avoided by adding a total
derivative of the form $-\p(X F)$ giving a term $-F\p X$ in the action
and hence a $-\p X$ in the algebraic equation of motion for $F$.
The second case is the troublemaker. If a term like $\p F \p F$
occurs, then a simple total derivative added to the action cannot mend 
the problem.
If the second type of terms does not appear, then a supersymmetric 
theory without the auxiliary field problem can be constructed.
First examples of such constructions include
refs.~\cite{Karlhede:1987bg,Buchbinder:1994iw,Gates:1995fx,Nitta:2001rh}. 
The manifestly supersymmetric term found in
ref.~\cite{Buchbinder:1994iw} offers a manifestly supersymmetric class
of higher-derivative theories -- free from the auxiliary field problem
-- and has been further studied recently in 
refs.~\cite{Khoury:2010gb,Adam:2011hj,Adam:2013awa,Nitta:2014pwa,Nitta:2014fca,Nitta:2015uba}.  
As is clear from the above discussion, one off-shell supersymmetric
Lagrangian gives two (or more) possibilities for Lagrangians in terms
of component fields.
The first possibility is to set the auxiliary field to zero ($F=0$),
which is possible if no superpotential is turned on; we call this the
canonical branch. 
The other possibility is to find a nontrivial solution for $F$ from
its equation of motion, which is algebraic because our construction
avoids the auxiliary field problem. We call this the non-canonical
branch.

This non-canonical branch is in fact key to constructing a
supersymmetric extension of the Skyrme term. 
It was found in ref.~\cite{Adam:2013awa} (see also
refs.~\cite{Nitta:2014pwa,Nitta:2015uba,Bolognesi:2014ova})
that this was the case for the baby Skyrme model, 
that is a 2+1 dimensional analog of the Skyrme model.
Namely, in this case, a supersymmetric baby Skyrme term 
free from the auxiliary field problem and four time derivatives 
was constructed, although the ordinary kinetic term cannot be 
included.
A BPS baby Skyrmion is of a compacton type \cite{Adam:2013awa} 
and is a 1/4 BPS state \cite{Nitta:2014pwa,Nitta:2015uba}
preserving a quarter of the original supersymmetry.

Higher derivatives usually come about in low-energy effective theories
by means of integrating out heavy states/fields. Often in the
low-energy limit of a theory, some global symmetries are broken
spontaneously by the vacuum. In this case the low-energy effective
theory is described by the Nambu-Goldstone (NG) bosons of the symmetry
breaking, dictated by the symmetries of the vacuum. More precisely,
the dynamics of the NG bosons corresponding to a spontaneous symmetry
breaking from $G$ to $H$, is described by a nonlinear sigma model
whose target space is given by the coset $G/H$ \cite{Coleman:1969sm}. 
A prime (and most famous) example is chiral symmetry breaking in QCD,
giving rise to the three pions.

When working with a supersymmetric theory, spontaneous symmetry
breaking of global symmetries can of course still take place. The
situation is, however, modified with respect the standard
case which we just mentioned above \cite{Bando:1983ab}.
Spontaneous symmetry breaking in a supersymmetric theory is tantamount
to a chiral superfield acquiring a vacuum expectation value (VEV) due
to the superpotential $W$ (F-term condition). The superpotential is a
holomorphic functional and hence the F-term condition giving rise to
the symmetry breaking is invariant under a larger group than
the original group $G$, namely the complexification of the group 
$G^{\mathbb{C}}$.
Now let us assume that the symmetry breaking again breaks $G$ to $H$.
As we already mentioned, the target space of the nonlinear sigma model
must be K\"ahler \cite{Zumino:1979et}. If the coset space $G/H$ is by
itself K\"ahler, then everything is as in the non-supersymmetric case
and all the NG bosons of the symmetry breaking are
genuine \cite{Itoh:1985ha}. 
However it often happens that the coset space $G/H$ is not
K\"ahler. In this case, supersymmetry enforces additional bosons to be 
massless; they are called quasi-NG bosons \cite{Bando:1983ab}.
It comes back to the complexification of the symmetry. A quasi-NG
always comes in a pair with a genuine NG; this is because the chiral
superfield is complex and the case of the quasi-NG arises when the
broken generator is Hermitian.
If for instance we consider an SU(2) subgroup, where the generators 
are the three Pauli matrices, an example of quasi-NG bosons is where 
the generator $\tau^3$ is broken, giving 1 genuine NG boson and 1
quasi-NG boson.
If on the other hand $\tau^1+i\tau^2$ is broken (which is
non-Hermitian), then we have 2 real genuine NGs. In the first case,
the multiplets corresponding to broken Hermitian generators are thus
called mixed or M-type, whereas those corresponding to non-Hermitian 
generators are called pure or P-type. 
Some conditions for the existence of quasi-NG bosons are known. 
(1) In the absence of gauge symmetry 
there must appear at least one M-type, and therefore one quasi-NG
boson \cite{Lerche:1983qa,Shore:1984bh}.  
(2) When the symmetry $G$ is broken by a VEV 
belonging to a real representation of $G$, 
there appear only M-type NGs leading to the same number of quasi-NG
bosons as the number of NG bosons, and  
the target space is $G^{\mathbb C}/H^{\mathbb C}\simeq T^*(G/H)$
\cite{Lerche:1983qa}. 
This is actually the case that we need for our purpose.
Namely, when chiral symmetry SU($N$)$_{\rm L}\times$ SU($N$)$_{\rm R}$
is spontaneously broken down to SU($N$)$_{\rm L+R}$,  
the target space is 
\beq
\frac{\mathrm{SL}(N,\mathbb{C})_{\rm L}
  \times\mathrm{SL}(N,\mathbb{C})_{\rm R}}
  {\mathrm{SL}(N,\mathbb{C})_{\rm L+R}}
\simeq
\mathrm{SL}(N,\mathbb{C}),
\eeq
which is the complexification of SU($N$).
The most general effective Lagrangian in this case 
was constructed in ref.~\cite{Kotcheff:1988ji}
and possible higher-derivative terms were constructed on the canonical 
branch \cite{Nitta:2014fca}.

In this paper, we construct a manifestly supersymmetric Lagrangian 
in the off-shell superfield formalism, based on
an SL($N$,$\mathbb{C}$)-valued field. 
We find that on the canonical branch it 
gives a Lagrangian with 2 and 4 derivatives. The fourth-order
derivative term is however not the Skyrme term; not even when the
field is restricted to an SU($N$) submanifold. 
On the non-canonical branch the situation is different, 
for which we will work with SU(2).
We are able to write down a manifestly supersymmetric Skyrme-like
Lagrangian density in terms of the SL(2,$\mathbb{C}$)-valued field,
which however does not take the simple form of the Skyrme term.
Nevertheless, if we restrict to a submanifold where the field only
takes values in SU(2), then the fourth-order derivative term reduces
exactly to the standard Skyrme term.
The quasi-NG bosons therefore change the Lagrangian density (and the
target space manifold) compared with the non-supersymmetric case.
But once they are turned off, the Skyrme term remains. 
One peculiarity of the solution on the non-canonical branch, however,
is that when turning on a kinetic term (the Dirichlet term), the
algebraic solution to the auxiliary field eliminates the kinetic term
and leaves only a potential behind. This is, however, a manifestly
supersymmetric way to introduce a potential in the ``extreme'' Skyrme
model without adding a superpotential. 
This was also the case for supersymmetric baby Skyrmions
\cite{Adam:2011hj,Nitta:2014pwa,Nitta:2015uba}. 
Our case is actually the first example of a solution on the
non-canonical branch in a model with multi-component chiral
superfields.

Returning to the problem of four time derivatives. For both the
solution on the canonical branch and non-canonical branch, the
Lagrangian possesses four time derivatives. However, in the case of
the non-canonical branch, when the field is restricted to the SU(2)
submanifold, then the terms with four time derivatives cancel out.

The paper is organized as follows. We set the notation and review the
construction of manifestly supersymmetric higher-derivative terms in
sec.~\ref{sec:formalism}, both for generic fields and for fields
taking values in a coset. Then in sec.~\ref{sec:susyskterm} we present 
our calculation of a supersymmetric Skyrme term as well as the
addition of the kinetic term.
In sec.~\ref{sec:gauging} we illustrate the possibility of gauging a
global symmetry of the model.
Finally we conclude with a discussion and outlook in
sec.~\ref{sec:conclusion}.

\section{The formalism}\label{sec:formalism}

In this section we briefly review the construction of a supersymmetric
higher-derivative term. 
We consider a higher-derivative Lagrangian that consists of 
$\mathcal{N} = 1$ chiral superfields.
We basically follow the Wess-Bagger conventions 
\cite{Wess:1992cp}.

\subsection{General action}

The component expansion of the chiral superfield in the $x$-basis is
given by 
\beq
\Phi^i(x,\theta,\bar{\theta})
= \varphi^i
+ i\theta\sigma^\mu\bar{\theta}\p_\mu\varphi^i
+ \frac{1}{4}\theta^2\bar{\theta}^2\square\varphi^i
+ \theta^2 F^i
+ \cdots
\eeq
where $i = 1, \ldots, N$ labels the multiple chiral superfields and 
the ellipses denote terms involving fermions.
Here $\varphi_i$ are complex scalar fields and $F_i$ are auxiliary
fields. 
The sigma matrices are defined as
$\sigma^\mu\equiv(\mathbf{1}_2,\vec{\tau})$, where
$\vec{\tau}=(\tau^1,\tau^2,\tau^3)$ are the Pauli matrices. 
The supercovariant derivatives read
\beq
D_{\alpha} = \frac{\partial}{\partial \theta^{\alpha}} + i
 (\sigma^\mu)_{\alpha \dot{\alpha}} \bar{\theta}^{\dot{\alpha}}
 \partial_\mu, \quad 
\bar{D}_{\dot{\alpha}} = - \frac{\partial}{\partial
\bar{\theta}^{\dot{\alpha}}} - i \theta^{\alpha} (\sigma^\mu)_{\alpha
\dot{\alpha}} \partial_\mu.
\eeq
A special combination of chiral superfields yields a
higher-derivative Lagrangian density without the auxiliary field
problem, which was mentioned in the introduction and can be written
as 
\beq
\frac{1}{16} \int d^4 \theta \;
\Lambda_{ik\bar{j} \bar{l}}
(\Phi,
 \Phi^{\dagger}) 
D^{\alpha} \Phi^i
D_{\alpha} \Phi^k \bar{D}_{\dot{\alpha}} \Phi^{\dagger \bar j}
\bar{D}^{\dot{\alpha}} \Phi^{\dagger \bar{l}}.
\label{eq:hd}
\eeq
The tensor $\Lambda_{ik\bar{j}\bar{l}}$ can be regarded as a (2,2)
K\"ahler tensor, which is symmetric in the two holomorphic indices,
$i$ and $k$ and it is also symmetric in the two anti-holomorphic
indices $\bar{j}$ and $\bar{l}$ \cite{Nitta:2014fca}.
This can be seen from the fact that $D_\alpha\Phi^i$ transforms like a
vector under field redefinitions. 

The bosonic components of the term \eqref{eq:hd} are given by
\begin{align}
\Lambda_{ik\bar{j} \bar{l}} (\varphi, \bar{\varphi})
\left[
(\partial_{\mu} \varphi^i \partial^{\mu} \varphi^k) (\partial_{\nu}
 \bar{\varphi}^{\bar{j}} \partial^{\nu} \bar{\varphi}^{\bar{l}})
-
\partial_{\mu} \varphi^i F^k 
\partial^{\mu} \bar{\varphi}^{\bar{j}} \bar{F}^{\bar{l}} 
+ F^i \bar{F}^{\bar{j}} F^k \bar{F}^{\bar{l}}
\right].
\end{align}
The auxiliary field with spacetime derivatives does not appear in
these terms\footnote{There are auxiliary fields with spacetime
derivatives in the fermionic sector. A solution to the auxiliary field
equation that contains fermions is obtained order by order in the
fermionic fields.
In this paper we concentrate on the bosonic sector for which analytic
solutions to the auxiliary field equation are found.}.
Then the supersymmetric chiral model with the higher-derivative term is
given by 
\begin{align}
\mathcal{L} =& \ \int \! d^4 \theta \ K (\Phi^i, \Phi^{\dagger \bar{j}}) 
+ \frac{1}{16} \int \! d^4 \theta \ 
\Lambda_{ik\bar{j} \bar{l}}
(\Phi, \Phi^{\dagger}) 
D^{\alpha} \Phi^i
 D_{\alpha} \Phi^k \bar{D}_{\dot{\alpha}} \Phi^{\dagger \bar j}
 \bar{D}^{\dot{\alpha}} \Phi^{\dagger \bar{l}} 
\notag \\
& + \left(\int \! d^2 \theta \ W(\Phi^i) + {\rm h.c.}\right)
\label{eq:hdLagrangian}
\end{align}
where $K$ is the K\"ahler potential and $W$ is a superpotential.
The equation of motion for the auxiliary field is 
\begin{align}
\frac{\partial^2 K}{\partial \varphi^i \partial \bar{\varphi}^{\bar{j}}} 
F^i - 2 \partial_{\mu} \varphi^i F^k \Lambda_{ik\bar{j} \bar{l}} 
 \partial^{\mu} \bar{\varphi}^{\bar{l}} + 
2 \Lambda_{ik\bar{j} \bar{l}} F^i F^k \bar{F}^{\bar{l}} 
+ \frac{\partial \bar{W}}{\partial \bar{\varphi}^{\bar{j}}} 
= 0. 
\label{eq:aux_eom}
\end{align}
Using this equation, we can eliminate the auxiliary fields and calculate
the on-shell Lagrangian.
Since eq.~\eqref{eq:aux_eom} is an algebraic equation of cubic order,
it has several solutions and there are distinct branches of on-shell 
Lagrangians associated with these solutions \cite{Sasaki:2012ka}.
Throughout this paper, we consider the $W=0$ case for simplicity. 
In this case, $F_i=0$ is always a solution.
For this solution, the on-shell Lagrangian reads
\begin{align}
\mathcal{L}_b = - \frac{\partial^2 K}{\partial \varphi^i \partial
 \bar{\varphi}^{\bar{j}}} \partial_{\mu} \varphi^i \partial^{\mu}
 \bar{\varphi}^{\bar{j}} 
+ \Lambda_{i k \bar{j} \bar{l}} 
(\partial_{\mu} \varphi^i \partial^{\mu} \varphi^k) (\partial_{\nu} \bar{\varphi}^{\bar{j}}
 \partial^{\nu} \bar{\varphi}^{\bar{l}}).
\end{align}
We call this the canonical branch.
The Lagrangian contains the ordinary kinetic term and a fourth-order
derivative term. One can take the small $\Lambda$ limit on the
canonical branch. 
Therefore the higher-derivative interactions can be perturbatively 
introduced to the ordinary (second-order derivative) theory.
We note that since the K\"ahler tensor $\Lambda_{ik\bar{j}\bar{l}}$ is
an arbitrary function of the complex scalars $\varphi^i$, it can
indeed contain terms with arbitrary orders of spacetime derivatives. 
An example is the scalar part of the $\mathcal{N} = 1$ 
supersymmetric completion of the Dirac-Born-Infeld action for a single 
D3-brane \cite{Rocek:1997hi}. 
Other examples are the supersymmetric Faddeev-Skyrme model
\cite{Nitta:2014pwa}, the effective action of supersymmetric 
field theories \cite{Gates:1995fx, Antoniadis:2007xc, Gomes:2009ev,
Kuzenko:2014ypa}.

In general, there are more solutions than $F_i=0$; we call the branch
of solutions with $F_i\neq 0$ the non-canonical branch.
Even though the equation of motion for the auxiliary field is
algebraic, it is not so straightforward to find an analytic solution
for the non-canonical branch, since the equation is a simultaneous
equation of cubic power. 
For single chiral-superfield models, the analytic solutions to the
equation and the on-shell Lagrangian on the non-canonical branch has
been found in ref.~\cite{Nitta:2014pwa}.
For the single chiral-superfield case, the ordinary kinetic term 
cancels out on the non-canonical branch and the K\"ahler tensor
$\Lambda$ enters into the Lagrangian as $\Lambda^{-1}$.
Hence, one cannot take the small $\Lambda$ limit in the on-shell
Lagrangian.  
This means that the higher-derivative interactions are introduced
non-perturbatively on the non-canonical branch.

\subsection{Chiral symmetry breaking}

Now we will take the above considerations and apply them to the case
of a chiral symmetry breaking of the form
\beq
G = \mathrm{SU}(N)_{\rm L}\times \mathrm{SU}(N)_{\rm R} \to
H = \mathrm{SU}(N)_{{\rm L}+{\rm R}},
\eeq
giving the corresponding coset
\beq
G/H = \frac{\mathrm{SU}(N)_{\rm L}\times \mathrm{SU}(N)_{\rm
R}}{\mathrm{SU}(N)_{{\rm L}+{\rm R}}}\simeq \mathrm{SU}(N),
\eeq
which is spanned by NG modes. 
As we mentioned in the introduction, in the supersymmetric case, the
group $G$ is complexified and the group $H$ is changed to the
so-called complex isotropy group, see e.g.~ref.~\cite{Nitta:2014fca}.
This means that the target space relevant for the supersymmetric
nonlinear sigma model describing the symmetry breaking is
\beq
G^{\mathbb{C}}/\hat{H}\simeq
\mathrm{SU}(N)^{\mathbb{C}} =
G^{\mathbb{C}}/H^{\mathbb{C}} \simeq
\mathrm{SL}(N,\mathbb{C}) \simeq
T^*\mathrm{SU}(N).
\eeq
The fact that $\hat{H}=H^{\mathbb{C}}$ here (which is not the case in
general), is because the realization of quasi-NG bosons is a so-called
maximal realization or fully-doubled realization \cite{Bando:1983ab,Lerche:1983qa}. 

We now represent the coset with the following nonlinear sigma model
field
\beq
M = \exp\left(i\Phi^A t^A\right)\in G^{\mathbb{C}}/\hat{H},
\eeq
where the NG superfields take the form
\beq
\Phi^A(y,\theta) =
\pi^A(y) + i\sigma^A(y)
+\theta\psi^A(y)
+\theta^2 F^A(y),
\eeq
where $t^A$ are the generators of $G^{\mathbb{C}}/\hat{H}$, $\pi^A$
are (genuine) NG bosons, $\sigma^A$ are quasi-NG bosons, $\psi^A$ are
quasi-NG fermions and finally, $F^A$ are auxiliary fields.

The NG supermultiplet obeys the following nonlinear transformation law 
\beq
M \to M' = g_{\rm L} M g_{\rm R}, \qquad
(g_{\rm L},g_{\rm R}) \in
\mathrm{SU}(N)_{\rm L}\times\mathrm{SU}(N)_{\rm R}.
\eeq
This yields the transformations
\beq
M M^\dag \to g_{\rm L} M M^\dag g_{\rm L}^\dag,
\eeq
implying that the K\"ahler potential can be constructed as a function
of $\Tr(M M^\dag)$.
The simplest example is
\beq
K_0 = f_\pi^2 \Tr(M M^\dag),
\label{eq:K0simplest}
\eeq
leading to the free bosonic Lagrangian density
\beq
\mathcal{L}_{0,b}^{(2)} = -f_\pi^2\Tr(\p_\mu M\p^\mu M^\dag),
\label{eq:L0b2}
\eeq
where $M$ in the last equation is the lowest component of the NG
multiplet with the same symbol. 
The K\"ahler potential \eqref{eq:K0simplest} is, however, not general;
the most general K\"ahler potential can be written as 
\cite{Kotcheff:1988ji} 
\beq
K = f\left[\Tr[M M^\dag],\Tr[(M M^\dag)^2],\cdots,\Tr[(M M^\dag)^{N-1}]\right],
\eeq
with an arbitrary functional of the $N-1$ variables.
Physically, the reason why the function can be arbitrary is due to the
existence of the quasi-NG bosons \cite{Shore:1988mn,Higashijima:1997ph}.
The isometry of the target space is $G$ and not its complexification
$G^{\mathbb{C}}$ and hence the target manifold is not homogeneous.
The shape of the manifold can be deformed along the directions of the
quasi-NG bosons, keeping the isometry $G$.
By setting the quasi-NG bosons to zero, we restrict to a submanifold
\beq
U = \left.M\right|_{\sigma^A=0}\in \mathrm{SU}(N),
\eeq
which in turn simplifies the kinetic term from the K\"ahler potential
to the usual chiral Lagrangian
\beq
\left.\mathcal{L}_{0,b}^{(2)}\right|_{\sigma^A=0}
= -f_\pi^2\Tr(\p_\mu U \p^\mu U^\dag),
\eeq
where $f_\pi$ and $f$ are related.

\section{The supersymmetric Skyrme term}\label{sec:susyskterm}

In this section we consider a fourth-order derivative term 
based on the formalism presented in the last section.

\subsection{A fourth-order derivative term in the chiral Lagrangian} 

The fourth-order derivative term that we consider is of the form
\begin{align}
\mathcal{L}_0^{(4)} &=
\frac{1}{16}\int d^4\theta\; 
\Lambda_{ik\bar{j}\bar{l}}(\Phi,\Phi^\dag)
D^\alpha \Phi^i D_\alpha \Phi^k
\bar{D}_{\dot{\alpha}}\Phi^{\bar{j}\dag}\bar{D}^{\dot{\alpha}}\Phi^{\bar{l}\dag}
\non
&=
\frac{1}{16}
\int d^4\theta\; \Lambda(M,M^\dag) \Tr[
D^\alpha M\bar{D}_{\dot{\alpha}} M^\dag
D_\alpha M\bar{D}^{\dot{\alpha}} M^\dag],
\label{eq:L4}
\end{align}
which is the simplest candidate for a fourth-order derivative
term \cite{Nitta:2014fca} and $\Lambda_{ik\bar{j}\bar{l}}$ is a
$G$-covariant $(2,2)$ K\"ahler tensor determined from the right-hand 
side of the above equation. 
$\Lambda(M,M^\dag)$ is a $G$-invariant real scalar on 
the target space, given by
\begin{align}
\Lambda(M,M^\dag) = g\left[\Tr[M M^\dag],\Tr[(M M^\dag)^2],\cdots,\Tr[(M M^\dag)^{N-1}]\right],
\end{align}
with an arbitrary function $g$ of $N-1$ variables. 
It is also possible to consider a non-$G$-invariant function
for $\Lambda(M,M^\dag)$ such as
$\Lambda(M,M^\dag) = \Tr (M+M^\dag)$.

The term \eqref{eq:L4} is of course not the most general term with
four derivatives, avoiding the auxiliary field problem, that can be
written down, see e.g.~ref.~\cite{Nitta:2014fca} for more general
four-derivative terms.  
The bosonic part of the term reads
\beq
\mathcal{L}_{0,b}^{(4)} =
\Lambda(M,M^\dag)\Tr\left[
  M_\mu^\dag M_\nu M^{\mu\dag} M^\nu
  + (F^\dag F)^2
  - M_\mu^\dag M^\mu F^\dag F
  - M_\mu M^{\mu\dag} F F^\dag
\right],
\label{eq:L0b4}
\eeq
where we have introduced the notation $M_\mu\equiv\p_\mu M$, which we
will use throughout the paper.

\subsection{Canonical branch}

We can now construct a theory with second-order and fourth-order
derivative terms, for instance, by adding the two
terms \eqref{eq:L0b2} and \eqref{eq:L0b4}
\beq
\mathcal{L}_0 = \mathcal{L}_0^{(2)} + \mathcal{L}_0^{(4)}.
\eeq
As we do not consider adding superpotentials to the theory in this
paper, the simplest solution to $F$ is given by putting $F$ on the
canonical branch: $F=0$, giving
\beq
\mathcal{L}_{0,b} =
-f_\pi^2\Tr M_\mu M^{\mu\dag}
+ \Lambda(M,M^\dag)\Tr M_\mu^\dag M_\nu M^{\mu\dag} M^\nu.
\label{eq:Lcanonical}
\eeq
Comparing the bosonic part of this theory with the Skyrme model, the
fourth-order derivative term is clearly different from the Skyrme
term.
First of all, it is not a curvature term (i.e.~cannot be written as a
curvature of a tensor) and second of all, it contains four time
derivatives and hence suffers from the Ostrogradsky
instability \cite{Ostrogradsky}.
As we mentioned already in the introduction, this feature is typical
for supersymmetric higher-derivative theories.

\subsection{Non-canonical branch: a supersymmetric Skyrme term}
As is clear from the discussion, there is also the possibility of
considering the non-canonical branch of solutions for the auxiliary
fields, i.e.~$F\neq 0$. 
For non-Abelian valued superfields, no explicit solutions to the
auxiliary field equation has been found so far, even for single field 
models. 
A prototypical example of such a non-Abelian valued field is the
Skyrme field.
We will explicitly construct the 
first solution for models with non-Abelian valued fields.
Let us start with the case in the absence of the kinetic term.

The equation of motion for the
auxiliary field coming from the Lagrangian density \eqref{eq:L0b4} 
reads
\beq
2F F^\dag F - F M_\mu^\dag M^\mu - M_\mu M^{\mu\dag} F = 0,
\label{eq:varF}
\eeq
when varied with respect to $F^\dag$ and 
\beq
2 F^\dag F F^\dag - M_\mu^\dag M^\mu F^\dag - F^\dag M_\mu M^{\mu\dag}
= 0.
\label{eq:varFdag}
\eeq
when varied with respect to $F$, which is simply the complex conjugate
of eq.~\eqref{eq:varF}.
We have assumed that $\Lambda(M,M^\dag)\neq 0$, which multiplies the
whole equation (the $\Lambda(M,M^\dag)=0$ solution is just turning off
the fourth-order derivative term).


Multiplying \eqref{eq:varF} by $F^{-1}$ from the right and
\eqref{eq:varFdag} by $(F^{\dag})^{-1}$ from the left gives us
\begin{align}
  2F F^\dag - F M_\mu^\dag M^\mu F^{-1} - M_\mu M^{\mu\dag} = 0,
  \label{eq:FFdag1} \\
  2F F^\dag - (F^\dag)^{-1} M_\mu^\dag M^\mu F^\dag - M_\mu
  M^{\mu\dag} = 0.
  \label{eq:FFdag2}
\end{align}
Subtracting \eqref{eq:FFdag2} from \eqref{eq:FFdag1}, we get
\beq
\left[M_\mu^\dag M^\mu, F^\dag F\right] = 0.
\label{eq:commute1}
\eeq
This means that they are diagonalizable in the same basis.

Now we do almost the same, but for $F^\dag F$.
Multiplying \eqref{eq:varF} by $F^{-1}$ from the left and
\eqref{eq:varFdag} by $(F^\dag)^{-1}$ from the right gives us now
\begin{align}
  2F^\dag F - M_\mu^\dag M^\mu - F^{-1} M_\mu M^{\mu\dag} F = 0,
  \label{eq:FdagF1} \\
  2F^\dag F - M_\mu^\dag M^\mu - F^\dag M_\mu M^{\mu\dag}
  (F^\dag)^{-1} = 0.
  \label{eq:FdagF2}
\end{align}
Subtracting now \eqref{eq:FdagF2} from \eqref{eq:FdagF1}, we get
\beq
\left[M_\mu M^{\mu\dag}, F F^\dag\right] = 0.
\label{eq:commute2}
\eeq
This means that these two matrices are also diagonalizable in the same
basis.

A Hermitian matrix $W$ can be written as
\beq
W = U D U^\dag,
\eeq
where $U$ is a special unitary matrix and $D$ is a diagonal matrix
composed of the real eigenvalues of $W$ in its diagonal.

We will now write the auxiliary field as
\beq
F = A R B,
\label{eq:FARB}
\eeq
where as we will see shortly, $R$ is a diagonal matrix except for when
some of the eigenvalues are degenerate, and $A$ is a special unitary
matrix that diagonalizes $F F^\dag$: 
\beq
F F^\dag = A R R^\dag A^\dag,
\eeq
while $B$ is a special unitary matrix that diagonalizes $F^\dag F$: 
\beq
F^\dag F = B^\dag R^\dag R B.
\eeq
This means that $R R^\dag$ and $R^\dag R$ are both diagonal matrices.

The eigenvalues of both $F F^\dag$ and $F^\dag F$ must be the same due
to Sylvester's determinant theorem:
\beq
\det\left(\lambda\mathbf{1} - F^\dag F\right)
= \det\left(\lambda\mathbf{1} - F F^\dag\right),
\label{eq:characteristic_poly}
\eeq
but this does not prove that $R^\dag R = R R^\dag$ (that is, it
does not prove that the eigenvalues appear in the same order along the 
diagonal).\footnote{If both $A$ and $B$ are prepared as column vectors
of eigenvectors, then switching the order of some of the eigenvectors
in $A$ but not in $B$ renders $R^\dag R\neq R R^\dag$, even though
they are both diagonal matrices with the same real eigenvalues. }

Let us simplify the problem for the moment and consider $N=2$, namely
SL$(2,\mathbb{C})$.
We know from \eqref{eq:characteristic_poly} that the two eigenvalues
of $F^\dag F$ are equal to the two of $F F^\dag$. Both these matrices
are Hermitian and thus both eigenvalues are real (and positive
semi-definite, due to the fact that they are composed of a matrix
multiplied by its Hermitian conjugate). 
Let us set
\beq
R =
\begin{pmatrix}
  \alpha & \beta\\
  \gamma & \delta
\end{pmatrix}, \qquad
\alpha,\beta,\gamma,\delta\in\mathbb{C}, 
\eeq
and require that $R R^\dag$ is diagonal. That translates to the
following condition
\beq
\bar{\beta} = -\frac{\gamma\bar{\alpha}}{\delta},
\label{eq:sl2con1}
\eeq
while the condition that $R^\dag R$ is diagonal translates to
\beq
\bar{\beta} = -\frac{\gamma\bar{\delta}}{\alpha},
\label{eq:sl2con2}
\eeq
both constraints are happily satisfied with $\gamma=\beta=0$ and thus
$R$ is itself a diagonal matrix. In this case, it is obvious that
$R R^\dag = R^\dag R$ and the eigenvalues must be positive.

Let us now assume that $\beta\neq 0$ which means in turn that
$\gamma\neq 0$.
The two constraints \eqref{eq:sl2con1} and \eqref{eq:sl2con2} can be
solved by
\beq
R =
\begin{pmatrix}
  c e^{i\eta} & -\bar{\gamma} e^{i(\eta+\theta)}\\
  \gamma & c e^{i\theta}
\end{pmatrix}, \qquad
\gamma\in\mathbb{C}^*, \qquad
c,\eta,\theta\in\mathbb{R}.
\eeq
Now since $\det F=1=\det A\det R\det B=\det R$, we have
\beq
\det R = c^2 e^{i(\eta+\theta)} + |\gamma|^2 e^{i(\eta+\theta)} = 1,
\eeq
which is solved by $c=\sqrt{1-|\gamma|^2}$ and $\theta=-\eta$.
We now have
\beq
R = 
\begin{pmatrix}
  \sqrt{1-|\gamma|^2} e^{i\eta} & -\bar{\gamma}\\
  \gamma & \sqrt{1-|\gamma|^2} e^{-i\eta}
\end{pmatrix}, \qquad
\gamma\in\mathbb{C}^*,\; 0<|\gamma|<1, \qquad
\eta\in\mathbb{R}.
\label{eq:Rnotdiagonal}
\eeq
Now we have
\beq
R^\dag R = R R^\dag = \mathbf{1}_2,
\eeq
which means that again $R^\dag R=R R^\dag$ (but now they both
additionally have degenerate eigenvalues).
We do not have a proof for $N>2$, but let us consider $N=2$ for now. 

Let us now consider \eqref{eq:FFdag1} and \eqref{eq:FFdag2} with
\eqref{eq:FARB} which read 
\begin{align}
2R R^\dag
- R B M_\mu^\dag M^\mu B^\dag R^{-1}
- A^\dag M_\mu M^{\mu\dag} A = 0,
\label{eq:RRdag1} \\
2R R^\dag
- (R^\dag)^{-1} B M_\mu^\dag M^\mu B^\dag R^\dag 
- A^\dag M_\mu M^{\mu\dag} A = 0.
\label{eq:RRdag2}
\end{align}
If $R$ is diagonal, then the above equations simplify as
\beq
R B M_\mu^\dag M^\mu B^\dag R^{-1} = B M_\mu^\dag M^\mu B^\dag,
\label{eq:simpl1}
\eeq
and
\beq 
(R^\dag)^{-1} B M_\mu^\dag M^\mu B^\dag R^\dag
= B M_\mu^\dag M^\mu B^\dag.
\label{eq:simpl2}
\eeq

However, we should first consider the case that $R$ is \emph{not} a
diagonal matrix, which is the case where it takes the form
\eqref{eq:Rnotdiagonal}.

Let us consider eq.~\eqref{eq:RRdag1}, which now reads
\begin{align}
& 2\mathbf{1}_2 -
\begin{pmatrix}
  \sqrt{1-|\gamma|^2} e^{i\eta} & -\bar{\gamma}\\
  \gamma & \sqrt{1-|\gamma|^2} e^{-i\eta}
\end{pmatrix}
\begin{pmatrix}
  m_1 & \\
  & m_2
\end{pmatrix}
\begin{pmatrix}
  \sqrt{1-|\gamma|^2} e^{-i\eta} & \bar{\gamma}\\
  -\gamma & \sqrt{1-|\gamma|^2} e^{i\eta}
\end{pmatrix} \\
&\qquad
-
\begin{pmatrix}
  \tilde{m}_1 & \\
  & \tilde{m}_2
\end{pmatrix}
= 
\begin{pmatrix}
  * & -\bar{\gamma}\sqrt{1-|\gamma|^2}(m_1 - m_2) e^{i\eta}\\
  -\gamma\sqrt{1-|\gamma|^2}(m_1 - m_2)e^{-i\eta} & *
\end{pmatrix} = 0. \nonumber
\end{align}
There are now three possibilities. 
If $m_1=m_2$, then $B M_\mu^\dag M^\mu B^\dag$ is proportional to the
unit matrix $\mathbf{1}_2$ and so the simplification \eqref{eq:simpl1}
holds.
The two remaining options are $\gamma=0$ and $\gamma=e^{i\tau}$ with
$\tau\in\mathbb{R}$ a real phase.
In the first case, $R$ is diagonal and so the
simplification \eqref{eq:simpl1} holds again.
If, however, $|\gamma|=1$ then the situation corresponds to switching
the two eigenvalues of $M_\mu^\dag M^\mu$ but not of $M_\mu
M^{\mu\dag}$; this can be done easily by switching the order of the
eigenvectors in $B$, but not in $A$. Thus again we can use the
simplification \eqref{eq:simpl1}.
Using instead eq.~\eqref{eq:FdagF1}, we can show the analogous case
for $\tilde{m}_1$ and $\tilde{m}_2$.
Therefore the simplification \eqref{eq:simpl2} holds as well.

Since we have proven that the simplifications \eqref{eq:simpl1} and
\eqref{eq:simpl2} hold, we can now write
\begin{align}
2R R^\dag
- B M_\mu^\dag M^\mu B^\dag
- A^\dag M_\mu M^{\mu\dag} A = 0,
\label{eq:RRdag3} \\
2R^\dag R - B M_\mu^\dag M^\mu B^\dag
- A^\dag M_\mu M^{\mu\dag} A = 0,
\label{eq:RdagR3}
\end{align}
or equivalently
\begin{align}
2F F^\dag
- A B M_\mu^\dag M^\mu B^\dag A^\dag
- M_\mu M^{\mu\dag} = 0.
\label{eq:FFdag3} \\
2F^\dag F
- M_\mu^\dag M^\mu
- B^\dag A^\dag M_\mu M^{\mu\dag} A B = 0.
\label{eq:FdagF3}
\end{align}
Eqs.~\eqref{eq:RRdag3} and \eqref{eq:RdagR3} clearly show that
$R R^\dag = R^\dag R$.
Eq.~\eqref{eq:FFdag3} shows that $F F^\dag$ is equal to the average
of $M_\mu M^{\mu\dag}$ and $M_\mu^\dag M^\mu$ rotated twice (by two
consecutive special unitary transformations).
Similarly for $F^\dag F$.  

Inserting \eqref{eq:FFdag3} and \eqref{eq:FdagF3}
into \eqref{eq:L0b4}, we get
\begin{align}
\mathcal{L}_{0,b}^{(4)} =
\Lambda(M,M^\dag) \Tr\bigg[ &
  M_\mu^\dag M_\nu M^{\mu\dag} M^\nu
  -\frac{1}{4}M_\mu^\dag M^\mu M_\nu^\dag M^\nu
  -\frac{1}{4}M_\mu M^{\mu\dag} M_\nu M^{\nu\dag} \non &
  -\frac{1}{2}A^\dag M_\mu M^{\mu\dag} A B M_\nu^\dag M^\nu B^\dag 
  \bigg].
\label{eq:L2}
\end{align}
Note that both $A^\dag M_\mu M^{\mu\dag} A$ and
$B M_\nu^\dag M^\nu B^\dag$ are diagonal matrices and that
\beq
\Tr\left[A^\dag M_\mu M^{\mu\dag} A B M_\nu^\dag M^\nu B^\dag\right]
= \Tr
\begin{pmatrix}
  m_1 & \\
  & m_2
\end{pmatrix}
\begin{pmatrix}
  \tilde{m}_1 & \\
  & \tilde{m}_2
\end{pmatrix}
= m_1 \tilde{m}_1 + m_2 \tilde{m}_2.
\eeq
For SL(2,$\mathbb{C}$), it is possible to write the eigenvalue
equation as
\begin{align}
0&=\lambda^2 - \lambda\Tr[A^\dag M_\mu M^{\mu\dag} A]
+ \det[A^\dag M_\mu M^{\mu\dag} A] \non
&= \lambda^2 - \lambda\Tr[M_\mu M^{\mu\dag}]
+ \det[M_\mu M^{\mu\dag}] \non
&= \lambda^2 - \lambda\Tr[M_\mu M^{\mu\dag}]
+ \frac{1}{2}\Tr[M_\mu M^{\mu\dag}]^2
- \frac{1}{2}\Tr[M_\mu M^{\mu\dag} M_\nu M^{\nu\dag}],
\end{align}
where we have used the Cayley-Hamilton theorem,
$\det A=\tfrac{1}{2}[(\Tr A)^2-\Tr A^2]$, in the last line. 
We can therefore write
\begin{align}
  &\Tr\left[A^\dag M_\mu M^{\mu\dag} A B M_\nu^\dag M^\nu
    B^\dag\right] =
  \frac{1}{2}\left(\Tr[M_\mu M^{\mu\dag}]\right)^2 \label{eq:lastterm}\\
  &\pm\sqrt{\left(
    \Tr[M_\mu^\dag M^\mu M_\nu^\dag M^\nu]
    -\frac{1}{2}\left(\Tr[M_\mu M^{\mu\dag}]\right)^2\right)
    \left(
    \Tr[M_\mu M^{\mu\dag} M_\nu M^{\nu\dag}]
    -\frac{1}{2}\left(\Tr[M_\mu M^{\mu\dag}]\right)^2\right)}.
  \nonumber
\end{align}
The sign ambiguity depends on the order of the eigenvalues
(i.e.~holding $m_1$ and $m_2$ fixed and switching $\tilde{m}_1$ and
$\tilde{m}_2$; this can be done as already mentioned by switching the 
eigenvectors in $A$ but not in $B$). 

Putting the pieces together, we can write down the Lagrangian density
for the full fourth-order derivative term as
\begin{align}
&\mathcal{L}_{0,b}^{(4)} = 
\frac{\Lambda(M,M^\dag)}{2} \bigg\{\Tr\left[ 
  2M_\mu^\dag M_\nu M^{\mu\dag} M^\nu
  -\frac{1}{2}M_\mu^\dag M^\mu M_\nu^\dag M^\nu
  -\frac{1}{2}M_\mu M^{\mu\dag} M_\nu M^{\nu\dag}\right]
  \label{eq:Lexplicit}\\
  &-\frac{1}{2}\left(\Tr[M_\mu M^{\mu\dag}]\right)^2\non
  &\mp\sqrt{\left(
    \Tr[M_\mu^\dag M^\mu M_\nu^\dag M^\nu]
    -\frac{1}{2}\left(\Tr[M_\mu M^{\mu\dag}]\right)^2\right)
    \left(
    \Tr[M_\mu M^{\mu\dag} M_\nu M^{\nu\dag}]
    -\frac{1}{2}\left(\Tr[M_\mu M^{\mu\dag}]\right)^2\right)}
    \bigg\}. \nonumber
\end{align}

Let us consider the four time derivatives term, which is present in
the Lagrangian density. For convenience, we consider the form of the
Lagrangian density in eq.~\eqref{eq:L2} and thus we have
\beq
\frac{1}{2}\Lambda(M,M^\dag) \Tr\left[ 
  M_0^\dag M^0 \left(M_0^\dag M^0
    - B^\dag A^\dag M_0 M^{0\dag} A B\right)
  \right],
\eeq
which means that the term does not cancel out due to the mismatch
between $M_\mu^\dag M^\mu$ and $M_\mu M^{\mu\dag}$ being
diagonalizable in the same basis ($A$ is generically not equal to
$B^\dag$).
One would expect this term to be a small four time
derivative term compared with that coming from the canonical branch,
see eq.~\eqref{eq:Lcanonical}.
Nevertheless, the above term is present in the Lagrangian and thus
gives rise to the Ostrogradsky instability \cite{Ostrogradsky}.

Now we want to show that we can reduce the Lagrangian density by
restricting the field $M\in$ SL$(2,\mathbb{C})$ to $U\in$ SU(2),
i.e.~a unitary field; that is, $M=U$ for which we now have
$U^\dag U=U U^\dag=\mathbf{1}_2$ and so
\beq
U_\mu^\dag U = -U^\dag U_\mu.
\eeq
This physically means that we turned off the quasi-NG bosons and
consider the submanifold spanned only by the genuine NG bosons. 
We first want to show that the eigenvalues of $U_\mu^\dag U^\mu$ are
the same as those of $U_\mu U^{\mu\dag}$. Hence, we have
\begin{align}
  &\det[\lambda\mathbf{1}_2 - U_\mu^\dag U^\mu] \non
  &=\det[\lambda\mathbf{1}_2 - U_\mu^\dag U U^\dag U^\mu] \non
  &=\det[\lambda\mathbf{1}_2 - U^\dag U_\mu U^{\mu\dag} U] \non
  &=\det\left[U^\dag(\lambda\mathbf{1}_2 - U_\mu U^{\mu\dag}) U\right]
  \non
  &=\det U^\dag \det[\lambda\mathbf{1}_2 - U_\mu U^{\mu\dag}]
  \det U \non
  &=\det[\lambda\mathbf{1}_2 - U_\mu U^{\mu\dag}].
\end{align}
This is a great simplification over the case with $M_\mu^\dag M^\mu$,
which in general does not have the same eigenvalues as
$M_\mu M^{\mu\dag}$.
Continuing, we can write the diagonal matrix
\beq
B U_\mu^\dag U^\mu B^\dag 
= B U_\mu^\dag U U^\dag U^\mu B^\dag 
= B U^\dag U_\mu U^{\mu\dag} U B^\dag,
\eeq
and thus we found a matrix that diagonalizes $U_\mu U^{\mu\dag}$,
namely $A = U B^\dag$.
We can now calculate the last term in eq.~\eqref{eq:L2} as
\beq
\Tr[A^\dag U_\mu U^{\mu\dag} A B U_\nu^\dag U^\nu B^\dag]
=\Tr[U^\dag U_\mu U^{\mu\dag} U U_\nu^\dag U^\nu]
=\Tr[U_\mu^\dag U^\mu U_\nu^\dag U^\nu]. \label{eq:Udiag}
\eeq
It is straightforward to show that
\begin{align}
  &\Tr[U_\mu U^{\mu\dag} U_\nu U^{\nu\dag}] \non
  &=\Tr[U_\mu U^\dag U U^{\mu\dag} U_\nu U^\dag U U^{\nu\dag}] \non
  &=\Tr[U U_\mu^\dag U^\mu U^\dag U U_\nu^\dag U^\nu U^\dag] \non
  &=\Tr[U_\mu^\dag U^\mu U_\nu^\dag U^\nu]. \label{eq:Uswitch}
\end{align}
Using now eqs.~\eqref{eq:Udiag} and \eqref{eq:Uswitch}, we can write
the bosonic Lagrangian density \eqref{eq:L2} for the restricted submanifold spanned
only by NG bosons, as
\beq
\mathcal{L}_{0,b}^{(4)} =
\Lambda(U,U^\dag) \Tr\left[
  U_\mu^\dag U_\nu U^{\mu\dag} U^\nu
  - U_\mu^\dag U^\mu U_\nu^\dag U^\nu
\right], \label{eq:LSkyrme}
\eeq
which for $\Lambda(U,U^\dag)={\rm const.}$ 
is exactly the Skyrme term.

We will now, for consistency, show that the same result follows from
the Lagrangian density \eqref{eq:Lexplicit}. It follows simply by
using eq.~\eqref{eq:Uswitch} and choosing the upper sign (which means
that the eigenvalues are not swapped), for which the last two terms of
the Lagrangian density combine to 
\beq
\frac{\Lambda(U,U^\dag)}{2}\Tr[U_\mu^\dag U^\mu U_\nu^\dag U^\nu],
\eeq
and the final result \eqref{eq:LSkyrme} again follows.

\subsection{The Dirichlet term}\label{sec:Dirichlet}

Now we add the normal kinetic term -- the Dirichlet energy -- to the
Lagrangian, which reads
\begin{align}
\mathcal{L} = f_\pi^2 \int d^4 \theta\; \Tr [M^{\dagger} M] 
+ \frac{1}{16} \int d^4 \theta\; \Lambda (M,M^{\dagger}) 
\Tr [\bar{D}_{\dot{\alpha}} M^{\dagger} D_{\alpha} M
 \bar{D}^{\dot{\alpha}} M^{\dagger} D^{\alpha} M].
\end{align}
The corresponding bosonic Lagrangian reads 
\begin{align}
\mathcal{L}_b &= f_\pi^2 \Tr 
[ 
- M^{\dagger}_{\mu} M^{\mu} + F^{\dagger} F 
]
\notag \\
&\phantom{=\ }
+ \Lambda (M,M^{\dagger}) \Tr 
[
M^{\dagger}_{\mu} M_{\nu} M^{\dagger \mu} M^{\nu} + (F^{\dagger} F)^2
- M^{\dagger}_{\mu} M^{\mu} F^{\dagger} F - M_{\mu} M^{\dagger \mu} F F^{\dagger}
].
\end{align}
giving rise to the equations of motion for the auxiliary fields
\begin{align}
f_{\pi}^2
\Lambda^{-1}(M,M^\dag) F + 2 F F^{\dagger} F - F M^{\dagger}_{\mu}
 M^{\mu} - M_{\mu}  M^{\dagger \mu} F = 0, 
\notag \\
f_{\pi}^2
\Lambda^{-1}(M,M^\dag) F^{\dagger} + 2 F^{\dagger} F F^{\dagger} -
 M^{\dagger}_{\mu} M^{\mu} F^{\dagger} - F^{\dagger} M_{\mu}
 M^{\dagger \mu} = 0. 
\label{eq:eomF+D}
\end{align}
For a trivial solution $F=0$, the Lagrangian on the canonical branch
is given in eq.~\eqref{eq:Lcanonical}.

Next, we study the non-canonical branch associated with an $F\neq 0$ 
solution. As in the case of the previous section, the equation
\eqref{eq:eomF+D} implies that $[M^{\dagger}_{\mu} M, F^{\dagger} F] = 0$.
Therefore $M^{\dagger}_{\mu} M^\mu$ and $F^{\dagger} F$ are again 
simultaneously diagonalizable. The same is true for
$M_{\mu} M^{\dagger \mu}$ and $F F^{\dagger}$. We can thus proceed
along the same lines as the discussion in the previous section, even
for the case where the ordinary kinetic term is turned on. 
The $F\neq 0$ solution for $G^{\mathbb{C}}=\mathrm{SL}(2,\mathbb{C})$
reads 
\begin{align}
& \ F F^{\dagger} = \frac{1}{2} 
\left[
AB M^{\dagger}_{\mu} M^{\mu} B^{\dagger} A^{\dagger} + M_{\mu}
 M^{\dagger \mu} - f_{\pi}^2 \Lambda^{-1} \mathbf{1}_2
\right], \notag \\
& \ F^{\dagger} F = \frac{1}{2} 
\left[
B^{\dagger} A^{\dagger} M_{\mu} M^{\dagger \mu} AB + M^{\dagger}_{\mu}
 M^{\mu} - f_{\pi}^2 \Lambda^{-1} \mathbf{1}_2
\right].
\end{align}
Eliminating the auxiliary fields by using the above solutions, we find
that the bosonic part of the Lagrangian on the non-canonical branch is 
given by 
\begin{align}
\mathcal{L}_b = \Lambda (M, M^{\dagger}) \Tr 
\bigg[&
M^{\dagger}_{\mu} M_{\nu} M^{\dagger \mu} M^{\nu} 
- \frac{1}{4} (M^{\dagger}_{\mu} M^{\mu})^2 - \frac{1}{4} (M_{\mu}
 M^{\dagger \mu})^2 \non
&
- \frac{1}{2} A^{\dagger} M_{\mu} M^{\dagger \mu} A
 B M^{\dagger}_{\nu} M^{\nu} B^{\dagger} 
- \frac{f_{\pi}^2}{4 \Lambda (M,M^{\dagger})} 
\mathbf{1}_2
\bigg].
\end{align}
We note that the ordinary kinetic term cancels out and only the 
fourth-order derivative term remains along with a potential-like term
which is proportional to $\Tr\mathbf{1}_2$; this term is absent when
$\Tr M^{\dagger} M$ term is not included. 
This is an alternative way
of introducing a potential term into supersymmetric theories without
the superpotential \cite{Koehn:2012ar}. 
The same structure has been found in the $N=1$
case \cite{Nitta:2014pwa}. 
The term $\Tr[A^\dag M_\mu M^{\dag\mu} A B M^\dag_\nu M^\nu B^\dag]$ is
evaluated as in section \ref{sec:susyskterm} and the first four terms 
can again be written explicitly as in eq.~\eqref{eq:Lexplicit}. 

We now restrict the field to $M = U \in \mathrm{SU}(2)$, as was done
in the previous section. The Lagrangian then reduces to
\begin{align}
\mathcal{L}_b |_{M=U} = \Lambda (U,U^{\dagger}) 
\Tr \left[
U^{\dagger}_{\mu} U_{\nu} U^{\dagger \mu} U^{\nu} -
 U^{\dagger}_{\mu} U^{\mu} U^{\dagger}_{\nu} U^{\nu}
\right] - \Tr 
\left[
\frac{f_{\pi}^2}{4 \Lambda (U,U^{\dagger})} \mathbf{1}_2
\right].
\end{align}
When we consider a $G$-invariant function $\Lambda$, 
the second term gives just a constant,
this is, simply the Skyrme term remains.
If we consider a $G$-variant function $\Lambda$
such as $\Lambda = \Tr (M + M^\dag)$,
the second term gives a potential term 
(but it also breaks the $G$-invariance of the Skyrme term).

\section{Gauging the global symmetry}\label{sec:gauging}

Finally, in this section, we point out that we can gauge a global
symmetry of the model and write down the interactions of the gauge
field.
We consider the $\text{SU}(N)_{\text{L}} \times \text{SU}(N)_{\text{R}}$ 
as a global symmetry of the Skyrme field $M$.
We introduce the $\mathcal{N} = 1$ vector multiplets $V_{\text{L}}$,
$V_{\text{R}}$ associated with SU$(N)_{\text{L}}$ and SU$(N)_{\text{R}}$
gauge groups respectively.
The vector multiplets are introduced in the gauge
covariantized supercovariant derivative which is defined by
\begin{align}
\mathcal{D}_{\alpha} M = D_{\alpha} M +
 \Gamma_{\alpha}^{\text{L}} M + M \Gamma^{\text{R}}_{\alpha},
\end{align}
where $\Gamma^{\text{L}, \text{R}}_{\alpha}$ are the superconnections given by
\begin{align}
\Gamma^{\text{L}}_{\alpha} = e^{- 2gV_{\text{L}}} D_{\alpha}
 e^{2gV_{\text{L}}}, \qquad 
\Gamma^{\text{R}}_{\alpha} = e^{2gV_{\text{R}}} D_{\alpha} e^{-2gV_{\text{R}}}.
\end{align}
Here $g$ is the gauge coupling constant.
Then the higher-derivative interaction becomes
\begin{align}
\frac{1}{16} \int \! d^4 \theta \ \Lambda (M,M^{\dagger},V_{\text{L}}, V_{\text{R}}) 
\Tr 
\left[
\bar{\mathcal{D}}_{\dot{\alpha}} M^{\dagger} 
e^{2gV_{\text{L}}}
 \mathcal{D}_{\alpha} M 
e^{-2gV_{\text{R}}} 
 \bar{\mathcal{D}}^{\dot{\alpha}} M^{\dagger} 
e^{2gV_{\text{L}}} 
 \mathcal{D}^{\alpha} M 
e^{-2gV_{\text{R}}}
\right].
\label{eq:gauged}
\end{align}
This term is invariant under the following $\text{SU} (N)_{\text{L}} \times
\text{SU}(N)_{\text{R}}$ gauge transformations
\begin{align}
e^{2gV_{\text{L,R}}} \ \longrightarrow \ e^{-i \lambda^{\dagger}_{\text{L,R}}} e^{2gV_{\text{L,R}}} e^{i
 \lambda_{\text{L,R}}}, \quad 
M \ \longrightarrow \ e^{-i \lambda_{\text{L}}} M e^{ i\lambda_{\text{R}}}, \quad 
M^{\dagger} \ \longrightarrow e^{-i \lambda^{\dagger}_{\text{R}}} M^{\dagger} e^{i \lambda^{\dagger}_{\text{L}}},
\end{align}
for a gauge invariant real function $\Lambda
(M,M^{\dagger},V_{\text{L}}, V_{\text{R}})$.
Here $\lambda_{\text{L,R}}$ are the $\text{SU}(N)_{\text{L}} \times
\text{SU}(N)_{\text{R}}$ chiral superfield gauge parameters.
It is straightforward to show that the bosonic component from
eq.~\eqref{eq:gauged} is given by 
\begin{align}
\Lambda (M,M^{\dagger},A^{\text{L,R}}_\mu) \Tr 
\bigg[&
D_{\mu} M^{\dagger} D_{\nu} M D^{\mu} M^{\dagger} D^{\nu} M
+ (F^{\dagger} F)^2 - D_{\mu} M^{\dagger} D^{\mu} M F^{\dagger} F \non
&- D_{\mu} M D^{\mu} M^{\dagger} F F^{\dagger} 
\bigg],
\label{eq:gauged_comp}
\end{align}
where $D_{\mu}$ is the gauge covariant derivative.
The term \eqref{eq:gauged_comp} possesses the same structure found in
the ungauged case. Therefore, we can trace the same procedure
discussed in sec.~\ref{sec:susyskterm} to find the solution of the
auxiliary field $F$. The action for the $N=2$ case is the same as the
one we found in sec.~\ref{sec:susyskterm}, with all the derivatives in  
the interactions replaced by the gauge covariant derivative. 
We can also introduce the ordinary gauge kinetic term in the Lagrangian
which would be relevant for the gauged BPS Skyrme
model \cite{Adam:2012pm}. 
The equation of motion for the auxiliary field $D$ in the vector
multiplet remains linear. We therefore integrate it out in a
trivial way.

\section{Conclusion and discussion}\label{sec:conclusion}

In this paper we have studied an $\mathcal{N} = 1$ supersymmetric
generalization of the Skyrme term in four spacetime dimensions. 
The model is based on the off-shell formulation of supersymmetric
higher-derivative theories free from the auxiliary field problem.
In order to find the Skyrme term, we consider
an SL($N$,$\mathbb{C}$)-valued chiral superfield $M$. Compared with
the single-component case, the equation of motion for the auxiliary
field, although algebraic, takes the form of a matrix equation and
hence it is not straightforward to eliminate the auxiliary fields. 
We have, however, shown that the equation can be explicitly solved for
the $N=2$ case, which is already a nontrivial case.
When the superpotential is absent, the trivial solution $F=0$ is
allowed. For this solution, the Lagrangian on the canonical branch
does not admit the Skyrme term. 
For the non-canonical branch, we have also found that the on-shell
Lagrangian generically depends on the double trace terms in addition
to the single trace terms. If we restrict the
SL(2,$\mathbb{C}$)-valued field $M$ to $M\in$ SU(2), then
the double-trace terms cancel out and leave behind only single trace
terms; the resulting Lagrangian is then nothing but the Skyrme term. 
The situation is essentially the same even when we introduce the
ordinary kinetic term 
$\int d^4\theta\Tr[M^\dag M]$.
We have found that the ordinary kinetic 
term $M_\mu^\dag M^\mu$ cancels out on the non-canonical branch just
as in the single-component case.
The Lagrangian on the non-canonical branch also admits the potential 
term (if we spoil the $G$-invariance of the Skyrme term)
which originates from the K\"ahler tensor $\Lambda$.

We note that it is possible to introduce a superpotential into the
model. 
In general, when one introduces a superpotential $W$, the equation of 
motion for the auxiliary field does not allow $F=0$ solution.
For single-component models, analytic solutions to the 
auxiliary field equation have been found \cite{Nitta:2014pwa}.
The superpotential induces quite non-linear higher-derivative
interactions in the on-shell action.
It would be interesting to find solutions for the auxiliary field in
our model with an added  superpotential and study the structure of 
the higher-derivative interactions induced by $W$.
Another interesting direction would be to study the supersymmetric
completion of our model with eight supercharges.

Another important issue is to study configurations that keep fractions
of supersymmetry in our model.
One of the characteristic features of the off-shell higher-derivative 
supersymmetric models is its good accessibility to BPS equations. 
The BPS equations are obtained by the condition that the supersymmetry 
transformation of the fermions vanish. The equations are completely 
determined by the solution of the auxiliary
field \cite{Nitta:2014pwa,Nitta:2015uba}. Therefore, our result is
quite important in the venue of finding BPS equations for Skyrme-like
models.  
Finding soliton solutions in our model requires further
investigations. 
We will come back to these issues in future studies.

The obvious generalization of our solution for the auxiliary field to
$N>2$ could also be interesting. $N=3$ would require some work, but is
probably doable. It would be interesting to see if a solution for
arbitrary large $N$ can be found. 

Finally, it would be interesting to study a manifestly supersymmetric
version of the BPS Skyrme model, namely a model with a sixth-order
derivative term.

\subsection*{Acknowledgments}

S.~B.~G.~thanks Roberto Auzzi for discussions. 
M.~N.~ thanks Koji Hashimoto for discussions at the early stage 
of this work.
S.~B.~G.~also thanks the Recruitment Program of High-end Foreign
Experts for support.
The work of M.~N.~is supported in part by a Grant-in-Aid for
Scientific Research on Innovative Areas ``Topological Materials
Science'' (KAKENHI Grant No.~15H05855) and ``Nuclear Matter in Neutron
Stars Investigated by Experiments and Astronomical Observations''
(KAKENHI Grant No.~15H00841) from the the Ministry of Education,
Culture, Sports, Science (MEXT) of Japan. The work of M.~N.~is also
supported in part by the Japan Society for the Promotion of Science
(JSPS) Grant-in-Aid for Scientific Research (KAKENHI Grant
No.~25400268) and by the MEXT-Supported Program for the Strategic
Research Foundation at Private Universities ``Topological Science''
(Grant No.~S1511006).
The work of S. S. is supported in part by Kitasato University
Research Grant for Young Researchers.

\end{document}